\documentclass[a4paper]{jpconf}
\usepackage{graphicx}
\usepackage{amsmath,amsthm, amssymb, latexsym}

\begin{document}
\title{Towards a FPGA-controlled deep phase modulation interferometer}

\author{M Ter\'an$^1$, V Mart\'in$^1$, Ll Gesa$ ^1$, I Mateos$^1$, F. Gibert$^1$, N. Karnesis$^1$, 
J Ramos-Castro$^2$, T S Schwarze$^3$, 
O Gerberding$^3$, G Heinzel$^3$, F~Guzm\'an$^4$ and M Nofrarias$^1$}

\address{$^1$ Institut de Ci\`encies de l'Espai, (CSIC-IEEC), Bellaterra, Spain}
\address{$^2$ Universitat Polit\`ecnica de Catalunya, Barcelona, Spain}
\address{$^3$ Albert-Einstein-Institut, Max-Planck-Institut f\"ur Gravitationsphysik, Hannover, Germany}
\address{$^4$ National Institute of Standards and Technology, Gaithersburg, MD 20899, USA}

\ead{nofrarias@ice.cat}

\begin{abstract}
Deep phase modulation interferometry was proposed as a method to enhance homodyne interferometers to work over many fringes. In this scheme, a sinusoidal phase modulation is applied in one arm while the demodulation takes place as a post-processing step. In this contribution we report on the development to implement this scheme in a fiber coupled interferometer controlled by means of a FPGA, which includes a LEON3 soft-core processor. The latter acts as a CPU and executes a custom made application to communicate with a host PC. In contrast to usual FPGA-based designs, this implementation allows a real-time fine tuning of the parameters involved in the setup, from the control to the post-processing parameters. 
\end{abstract}

\section{Introduction}
Deep phase modulation interferometry~\cite{Heinzel10} was proposed as a method to enhance homodyne interferometers to work over many fringes, allowing for instance continuous real-time tracking of a free falling test mass, as required for space based gravitational wave detectors~\cite{elisa}.
The advantage of the proposed deep phase modulation scheme is that simplifies the required optical setup, driving and modulation electronics when compared with heterodyne based detection experiments~\cite{Heinzel05}. The technique was proposed as an extension of the so called $J_1 \ldots J_4$~\cite{Sudarshanam} methods, which involve a sinusoidal phase modulation at a fixed frequency with modulation depths up to 5\,rad in one arm of the interferometer. These methods extract the encoded phase by means of analytical formulae to solve for the unknowns, among them the phase. Deep phase modulation implies the usage of a higher amount of harmonics which requires the use of numerical least-squares techniques, allowing an increase of the signal-to-noise ratio and consequently leading to a improved accuracy (up to $\rm 20\,pm/\sqrt{Hz}$ above 10\,mHz ~\cite{Heinzel10}) with a relatively simple setup. 

The deep phase modulation scheme uses however higher order harmonics ($n \ge 10$) to extract the phase information from the modulated output. For a phase modulated homodyne interferometer, this can be expressed as
\begin{equation}
V_{PD}(t) = V_{DC}(\phi) + \sum^{\infty}_{n=1} a_n(m, \phi) \cos(n\, (\omega_m\,t + \Psi)) 
\end{equation}
with
\begin{eqnarray}
a_n(m, \phi) & = & k\, J_n(m) \cos\left( \phi + n\, \frac{\pi}{2} \right)  \label{eq.an} \\
V_{DC}(\phi) & = & A (1 + C\, J_0(m)\, \cos \phi) 
\end{eqnarray}
where $J_n(m)$ are Bessel functions, $\phi$ is the interferometer phase, $m$ is the modulation depth, $\omega_m$ is the modulation frequency, $\Psi$ is the modulation phase, $C$ is the contrast, and $A$ combines nominally constant factors such as light powers and photodiode efficiencies. Hence, the proposed scheme requires an implementation that allows high index phase modulation and precise phase extraction, as we describe in the following.

\section{Experiment description and results}

\begin{figure}[t]
\begin{center}
        \includegraphics[width = 0.65 \columnwidth]{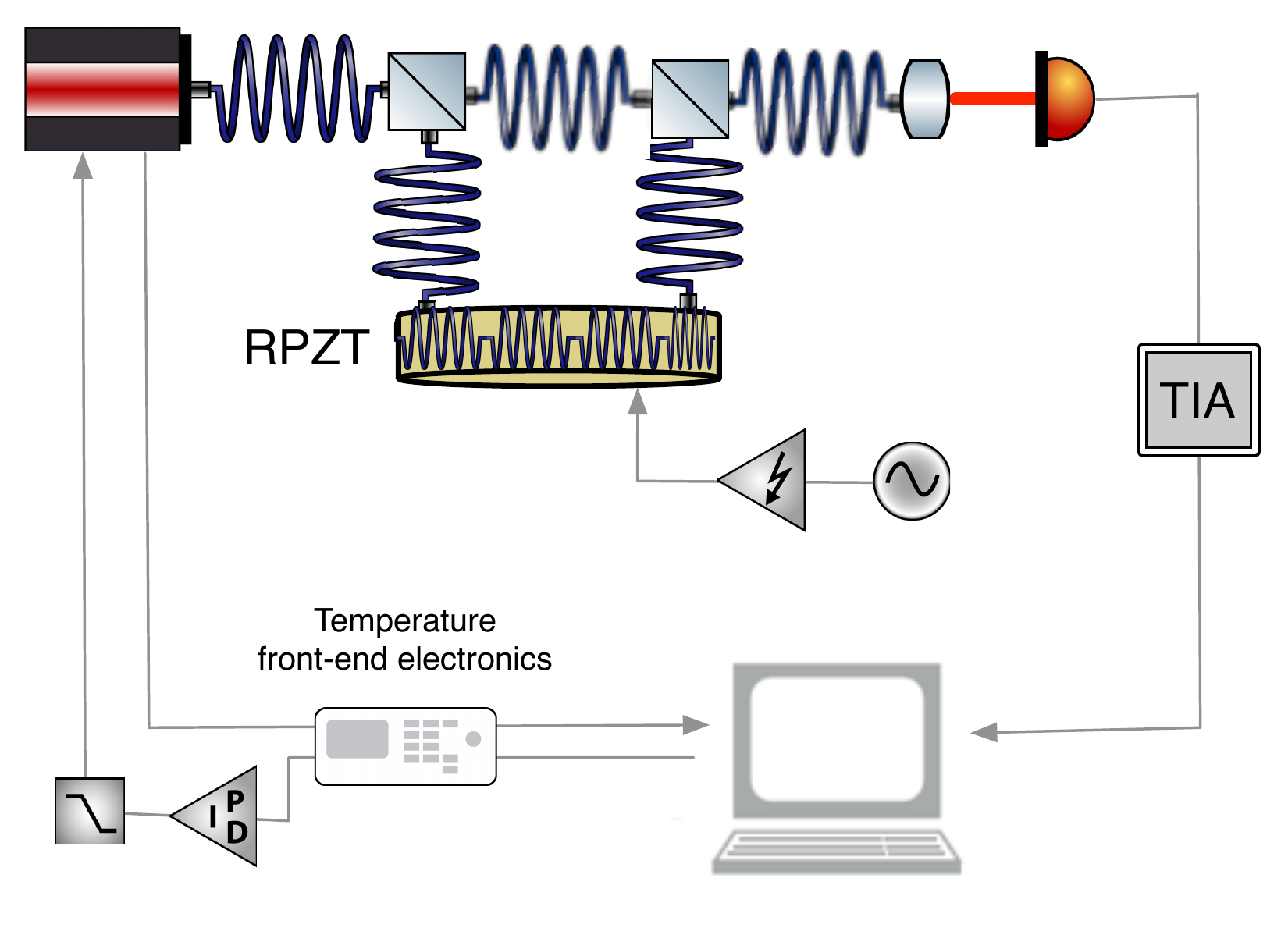}      
\caption{Deep phase modulation is implemented through an all-fiber Mach-Zehnder interferometer that uses a piezo tube with optical fiber coiled around it as a phase modulator. The measured output voltage is post processed in an external PC. \label{fig.scheme}}
\end{center}
\end{figure}

Our implementation of the deep phase interferometry is an all-fiber 
Mach-Zehnder interferometer operating at 1064\,nm which uses a piezo tube with 5\,m of optical fiber wrap around it. The modulation of the laser is achieved by increasing the path-length through a sinusoidal signal applied, after being amplified to high voltage, to the piezo tube. 
Although temperature stability is not currently limiting our setup we have implemented a temperature control loop setting the working temperature of the laser source. In our current setup, displayed in Fig.\ref{fig.scheme}, the LISA Pathfinder temperature front-end electronics~\cite{Sanjuan07} is modified to control the laser temperature by reading the thermistor in the laser diode and using this signal to set the temperature of the laser mount through a PID controlling the Peltier element in the laser. Data acquisition and post-processing is performed in a PC.

\begin{figure}[t]
\begin{center}
        \includegraphics[width = 0.49\columnwidth]{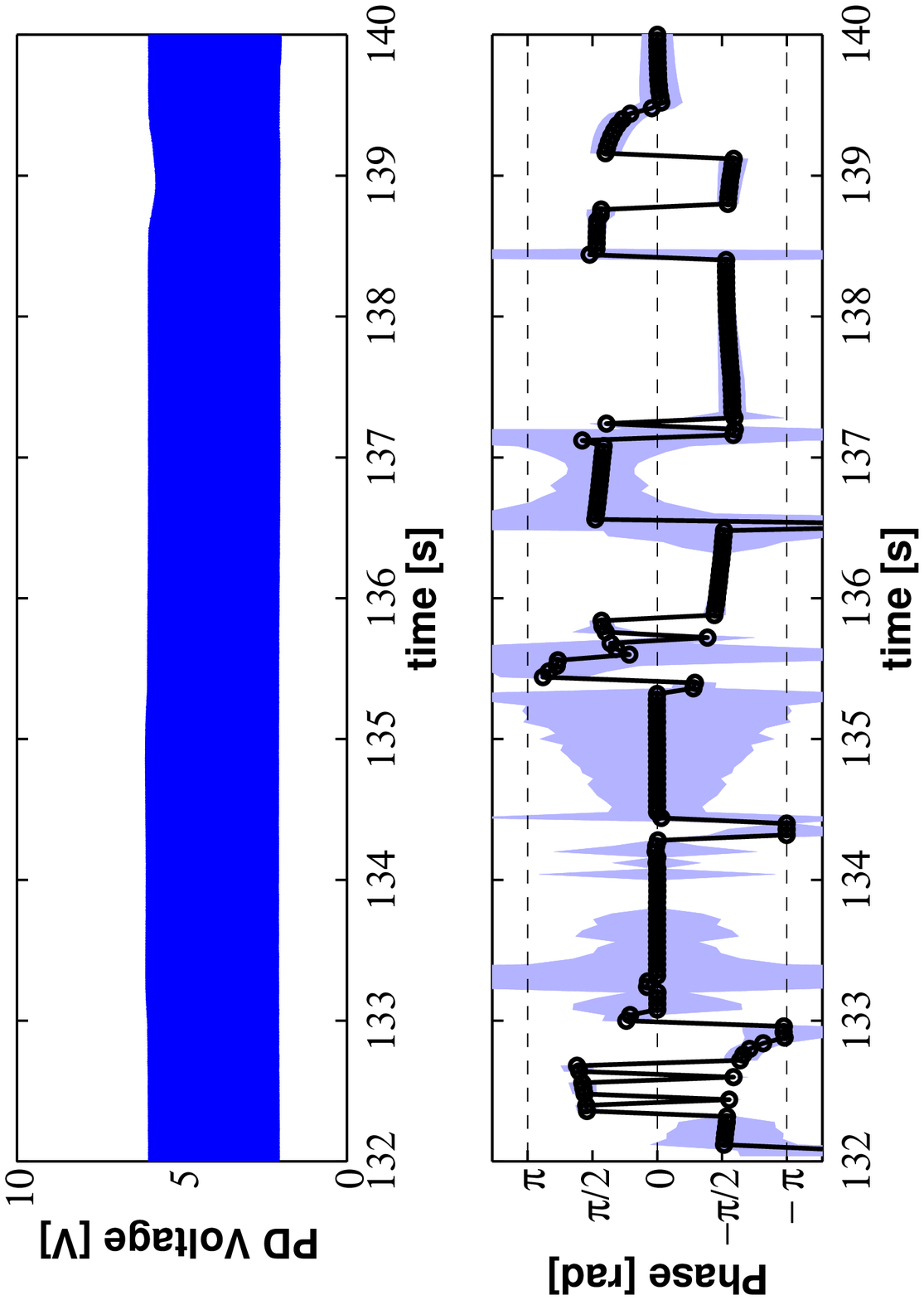}      
        \includegraphics[width = 0.5\columnwidth]{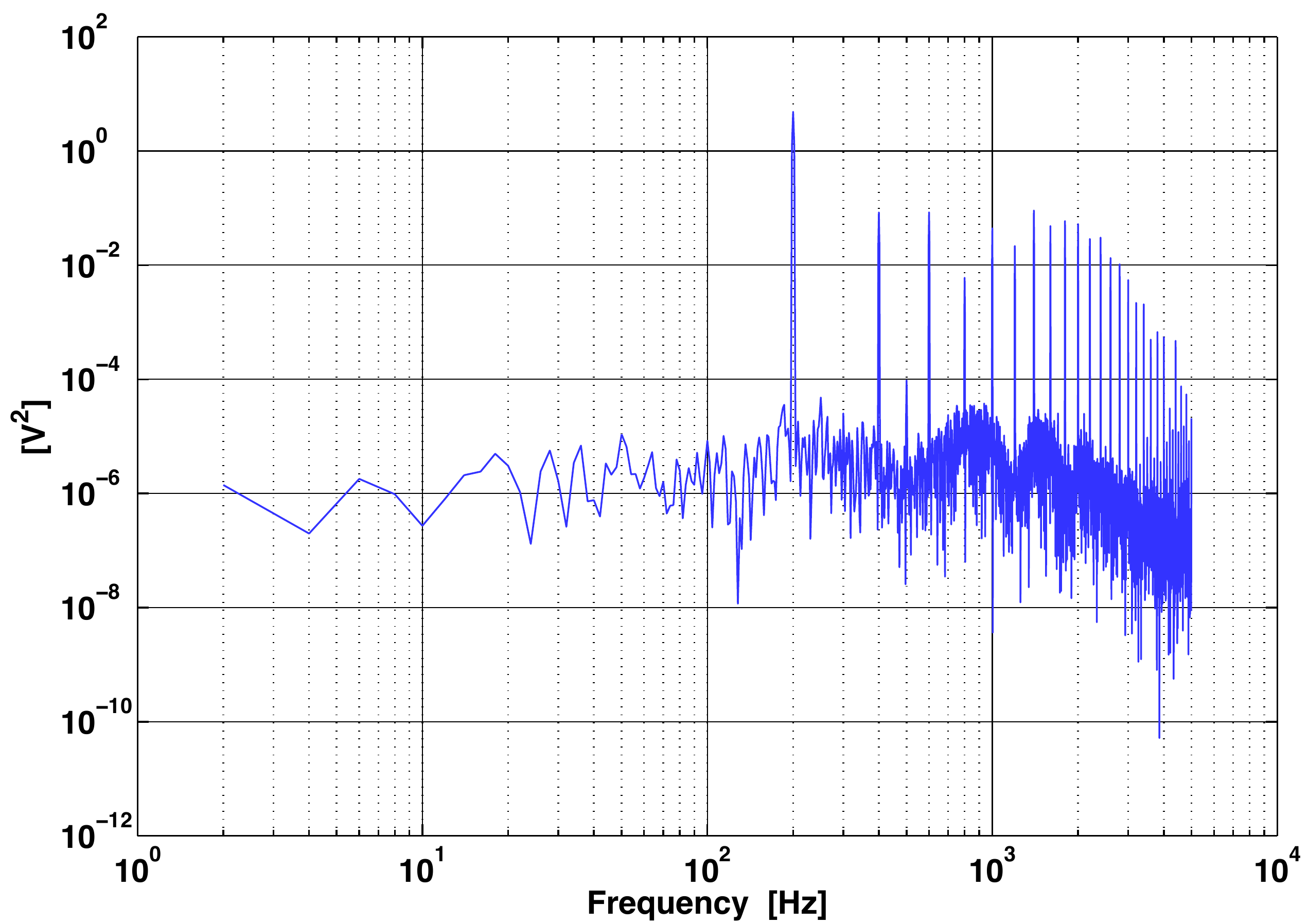}      
\caption{\emph{Left:} on top, the output voltage measured at the photodiode. Below, the measured phase after the post-processing. The shadowed area shows 95\,\% confidence intervals due to fit errors. \emph{Right:} A power spectrum of a single 4000 samples window showing the modulation pattern induced by the piezo. \label{fig.results}}
\end{center}
\end{figure}

In order to obtain the interferometer phase, the Fourier coefficients $a_n(m, \phi)$ in Eq.~(\ref{eq.an}) are first obtained through a Fast-Fourier Transform of a segment of data. 
Then, the coefficients $\lbrace k,m,\phi,\Psi \rbrace$  are obtained by minimisation of
\begin{equation}
\chi^2 = \sum^{10}_{n=1} \lvert \widetilde{V}_{PD}(n) -  a_n(m, \phi) e^{ i\,n\,\Psi} \rvert^2
\end{equation}
where $\widetilde{V}_{PD}(n)$ is the n-th harmonic of the measured voltage at the output of the photodiode. A Levenberg-Marquardt algorithm is used to process the measured output and obtain the set of coefficients.
In our proof-of-principle implementation we have applied a 200\,Hz modulation signal to the piezo with a modulation depth ${\rm m}\simeq 9$\,rad. Our current low frequency sensitivity with a table-top experiment on air is  $10\,{\rm \mu m/\sqrt{Hz}}$ at 10\,mHz which goes down to $10\,{\rm nm/\sqrt{Hz}}$ at 1\,Hz.
Figure \ref{fig.results} shows the photodiode output sampled at 10\,kHz and the associated phase after post-processing in windows of 4000 samples, yielding an effective phase sampling of 2.5\,Hz. The shaded areas show 95\,\% confidence intervals due to fit errors as derived from the errors obtained by the fit. As shown in Figure \ref{fig.results}, this interval allows the identification of areas where the errors provided by the fit are too big to provide conclusive values for the phase since the confidence interval spans a region greater than $2 \pi$. The origin of this error must be sought either in sudden changes in the voltages values being fit or due to inaccurate starting values for the non-linear fit.  
    
\section{The System On Chip approach}

While the original deep phase modulation interferometer was implemented in a scheme similar to the one shown in Fig.~\ref{fig.scheme}, recent developments have improved the design to shift the generation of signal modulation and the phase extraction to a FPGA~\cite{Schwarze14}. 

\begin{figure}[t]
\begin{center}
        \includegraphics[width = 0.65\columnwidth]{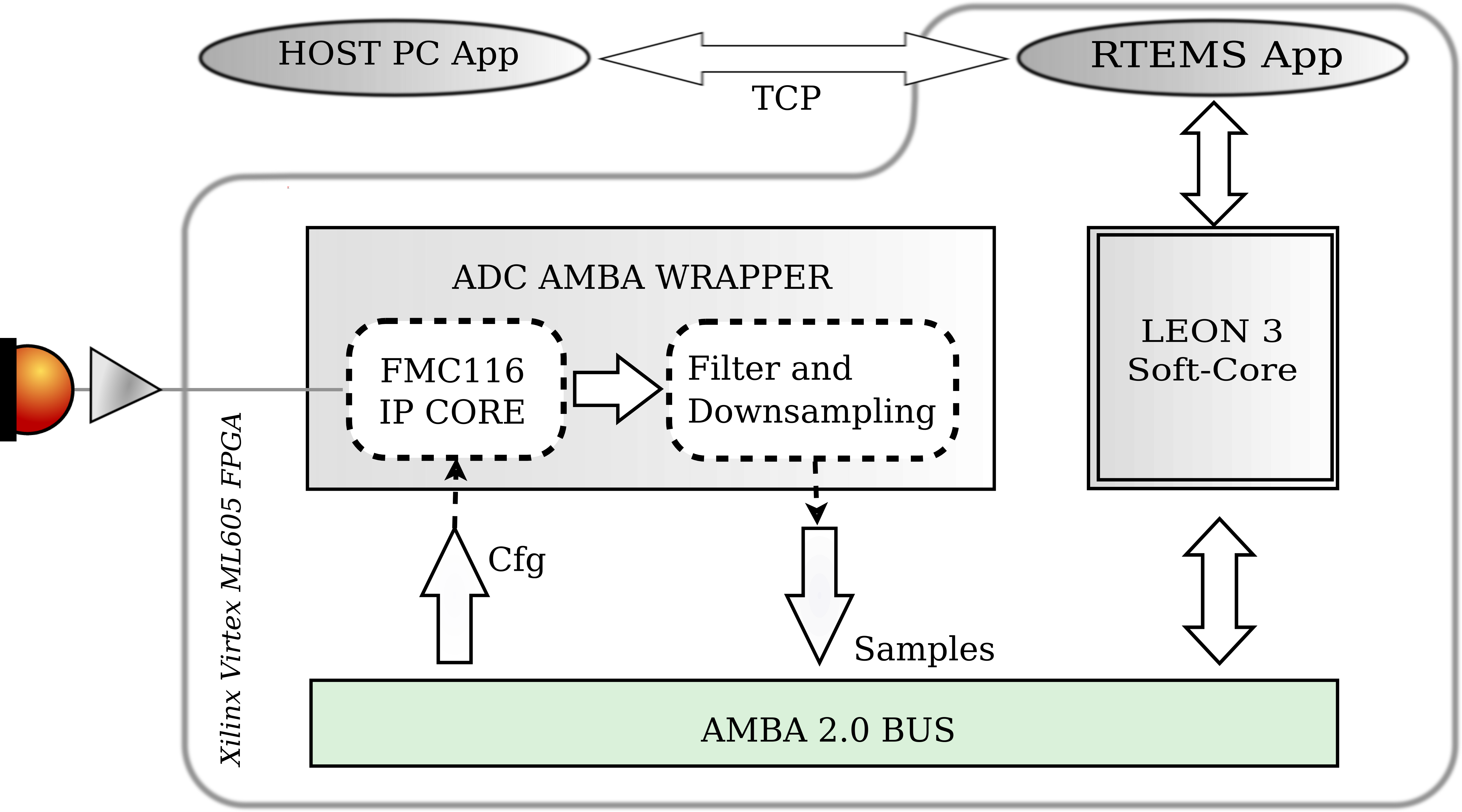}      
\caption{A schematic of the System On Chip (SoC) infrastucture in our current design. \label{fig.SOC}}
\end{center}
\end{figure}

In the same line, we have developed the software infrastructure that will allow a FPGA-based phasemeter configurable in real-time thanks to the System On Chip (SoC) approach. As shown in Fig~\ref{fig.SOC}, the following components have been synthesized in a Xilinx$^{\textcircled c}$ FPGA: A Gaisler$^{\small \textcircled c}$ LEON 3 Soft-Core CPU, a 4DSP$^{\textcircled c}$  FMC116 ADC wrapped in a custom made component that communicates directly with the CPU using AMBA technology bus, and a custom embedded RTEMS Application running on SoC, that is in charge of acquiring, processing and transmitting data to Host PC Application through ethernet TCP/IP, system monitoring and configuration managing. In parallel, the Host PC Application manages the user interface to customize the system and data persistence.

This infrastructure will allows us to implement the fit and parameter extraction step in the LEON 3 Soft-Core CPU, which will provide the already processed phase values to the Host PC Application.

\section{Conclusions}

We have shown an implementation of deep phase interferometer based on an all-fiber Mach-Zehnder interferometer with a piezo tube with fiber coiled around it acting as a phase modulator.  Our proof-of-concept experiment achieves a performance of  $10\,{\rm \mu m/\sqrt{Hz}}$ at 10\,mHz going down to $10\,{\rm nm/\sqrt{Hz}}$ at 1\,Hz. 

We have also implemented the required software infrastucture to run the required post processing tasks in a dedicated FPGA. Our effort has been focused so far in achieving a data acquisition system based on the SoC approach where we can build more complex data processing tasks at later stages. The LEON 3 Soft-Core CPU embedded in the FPGA will allow a more flexible control of the experiment once it is fully integrated in the optical setup.

Future developments of our setup include the already mentioned integration of the FPGA in the optical experiment, the implementation of the digital analysis, modulation and post-precessing in the FPGA and the integration of the metrology experiment in vacuum conditions.

\section*{Acknowledgements}
MN acknowledges a JAE-doc grant from CSIC and support from the EU Marie Curie CIG 322288.

\end{document}